\documentclass[fleqn,twoside]{article}
\usepackage{espcrc2}

\usepackage{graphicx}
\usepackage[figuresright]{rotating}

\newcommand{\AmS}{{\protect\the\textfont2
  A\kern-.1667em\lower.5ex\hbox{M}\kern-.125emS}}
\newcommand{\be}{\begin{equation}}
\newcommand{\ee}{\end{equation}}
\def\beqa{\begin{eqnarray}}
\def\enqa{\end{eqnarray}}

\def\nnb{\nonumber}
\def\Tr{\mbox{ Tr }}

\title{Tests of Universality of Baryon Form Factors in Holographic QCD}

\author{A. Cherman\address[MCSD]{Department of Physics, University of Maryland,
College Park, MD 20742-4111},
        T.D. Cohen\addressmark and
        M. Nielsen\address{Instituto de F\'{\i}sica, Universidade de 
S\~{a}o Paulo,
C.P. 66318, 05389-970 S\~{a}o Paulo, SP, Brazil}
        }
       
\begin{document}

\begin{abstract}
We describe a new exact relation for large $N_c$ QCD for the 
long-distance behavior of baryon form factors in the chiral limit, 
satisfied by all 4D semi-classical chiral soliton models.
We use this relation to test the consistency of the structure of 
two different  holographic models of baryons.
\vspace{1pc}
\end{abstract}

\maketitle

\section{Introduction}

There are still no systematic analytic tools to study the
strong-coupling dynamics of QCD, except for models that probe certain
limited classes of observables.  Thus, for example, one can use chiral
perturbation theory for some low-energy observables, but for more
general ones one is forced to resort to more phenomenological
approaches such as chiral soliton models.  In recent years,
holographic models of QCD have emerged as another approach
to the low energy phenomenology of QCD, and have attracted
considerable
interest \cite{EKSS,DRP1,KatzTensorMesons,KKSS,BrodskydeTeramond,tobias}.

There are two distinct classes of AdS/QCD models.  
Top-down AdS/QCD models arise from string theory, with the D4/D8 system
describing a gauge theory which is confining with non-Abelian chiral symmetry
breaking as in QCD. Despite the fact that the classical limit of the AdS/CFT
correspondence requires the number of colors and the 't Hooft coupling to be
 large \cite{Maldacena}, predictions of top-down AdS/QCD models, extrapolated
to three colors, fare relatively well when compared to experimental data
\cite{SS}. Botton-up model are also motivated by the AdS/CFT correspondence,
but are more phenomenological and allow more freedom to match QCD data
\cite{EKSS,DRP1,KKSS,BrodskydeTeramond}.
In these models, QCD in the large $N_c$ limit is taken to be dual to a
classical 5D theory in a curved space, and the parameters of the 5D
model are matched to their corresponding values in large $N_c$ QCD
\cite{tHooft,WittenNc},
with the field content of the 5D models chosen to match the low energy
chiral symmetry of QCD.  In contrast to approaches like chiral
perturbation theory, these models allow the computation of meson
spectra and couplings, at least in principle.  Even very simple 5D
models seem to show a remarkable phenomenological success when
compared to data.

Given the many bold assumptions that are necessary to construct
holographic models of QCD, their phenomenological success is
remarkable, and may suggest that the assumptions are more reliable
than might be expected.  It is natural to wonder if there is anything
in these models that can test the reliability of the assumptions. In this work
we use a new model-independent relation for baryons, to test two new 
holographic models of baryons \cite{PomarolWulzer,HSSY}. This relation
is sensitive to the anomalous coupling of the baryon current to three pions 
and becomes an exact result of QCD in the combined large $N_c$ and chiral 
limits (with the large $N_c$ limit taken first).

In the large distance limit $r \rightarrow \infty$, the ratio of 
position-space electric and magnetic baryon form factors is given by 
\cite{nos}:
\be
\label{Rre}
\lim_{r\rightarrow \infty} \frac{{G}_E^{I=0} {G}_E^{I=1}}{
{G}_M^{I=0} {G}_M^{I=1}} = \frac{18}{r^2} \;.
\ee

These position-space form factors can be related to the standard 
experimentally accessible momentum-space form factors by
\beqa
{G}_E^{I=0,1}(r) &= &\int{\frac{d^{3}q} {(2\pi)^3}~e^{i\vec{q}.\vec{r}}
\tilde{G}_E^{I=0,1}} (q)\nnb\\
{G}_M^{I=0,1}(r) &= &\frac{-i} {3}\int{\frac{d^{3}q} {(2\pi)^3}~e^{i
\vec{q}.\vec{r}}~\vec{q}} .\vec{r}~\tilde{G}_M^{I=0,1} (q).
\enqa

\section{5D Skyrmions}

The authors of refs. \cite{PomarolWulzer,PW} use a simple holographic model 
for QCD, in which the Chern-Simon (CS) term is incorporated to take into 
account the QCD 
chiral anomaly. They show that the baryons arise as stable solitons which are 
the 5D analog of the 4D skyrmions.  

The action of the 5D model is given by
\beqa
\label{s5d}
S &=& -\frac{M_5}{2} \int{d^{5}x \sqrt{g} \Tr[ {\cal L}^2_{MN}  + 
{\cal R}^2_{MN}]  }   \\
&+&{-i N_c\over24\pi^2} \int_{5D}{\left[\omega_5({\cal L})-\omega_5({\cal R})
\right]}\nnb,
\enqa
where ${\cal L}_{MN}, {\cal R}_{MN}$ are the U(2) gauge field strengths, 
$M,N = z,\mu $, $\omega_5$ is the Chern-Simons 5-form, and $M_5 \sim {\sc O}(
N_c^1)$ is an input parameter of the model.

In this model baryons are identified as the quantum states of slowly rotating 
`5D Skyrmions'.   The 5D Skyrmions are defined to be topologically 
non-trivial configurations of the 5D gauge fields with baryon number $B = 1$.
The hedgehog-like field configurations with $B=1$ can be parametrized in 
terms of five functions $\phi_{1}(r,z)$, $\phi_{2}(r,z)$, $A_{1}(r,z)$, 
$A_{2}(r,z)$, $s(r,z)$.  These functions satisfy 
some EOM, which are solved numerically.

By writing the isospin currents for the explicit case of a $B=1$, slowly 
rotating 5D Skyrmion, it is possible to show that
\beqa
G_E^{I=0}(r) &=& -{4\over N_c} M_5 \left[ a(z)\partial_z s\over r \right]_{z=0}
, \nnb\\
G_M^{I=0}(r)&=& -{2\over 3N_c\cal{I}} M_5 \left( r a(z)\partial_z Q~ 
\right)_{z=0}, \nnb\\
G_E^{I=1}(r) &=& {2\over 3\cal{I}}M_5 \left[a(z)\left(\partial_z v-2 
(\partial_z\chi_2\right.\right.\nnb\\
&-&\left.\left. A_2 \chi_1)\right) \right]_{z=0}, \nnb\\
G_M^{I=1}(r) &= &-{4\over9}M_5 \left[ a(z)(\partial_z \phi_2 - A_2 \phi_1)~
\right]_{z=0},
\label{cur5D}
\enqa
where $\cal{I}$  is the moment of inertia, and $v(r,z)$, $Q(r,z)$, $\chi_{1}
(r,z)$, $\chi_{2}(r,z)$ parametrize the collective rotations of the 5D 
Skyrmion, and are defined in ref.~\cite{PW}.

The authors of ref.~\cite{PW} also give the $r\to\infty$ behaviour of the
functions $s(r,z),~Q(r,z),~v(r,z),~\chi_i(r,z)~\phi_i(r,z),~A_i(r,z),~i=1,2$. 
These functions are parametrized in terms of the parameter
$\beta$ which is determined numerically. Using this we get for the large $r$
limit of the form factors:
\beqa
G_E^{I=0}(r\to\infty)&=&-{\beta^3L^6\over\pi^2}{1\over r^9},\nnb\\
G_M^{I=0}(r\to\infty)&=&{\beta^3L^6\over6\pi^2\lambda}{1\over r^7},\nnb\\
G_E^{I=1}(r\to\infty)&=&{8\beta^2\over3\lambda}M_5L^3{1\over r^4},\nnb\\
G_M^{I=1}(r\to\infty)&=&-{8\beta^2\over9}M_5L^3{1\over r^4},
\label{5dff}
\enqa

Using Eqs.~(\ref{5dff}) it is clear that Eq.~(\ref{Rre}) is satisfied
 in the Pomarol-Wulzer model of baryons as 5D Skyrmions.
This means that the Pomarol-Wulzer model correctly captures the large $N_c$ 
chiral physics of QCD to which Eq.~(\ref{Rre}) is sensitive.

\section{Baryons as Holographic Instantons}

In refs.~\cite{HSSY,HSS}, baryons are described as instantons in the 5D 
Yang-Mills (YM) and CS theory formulated in the D4/D8 model. 
It has been argued that the low energy phenomena of QCD can be derived from 
this model. In the case of the D4/D8 model, baryons are identified as D4-branes
wrapped on a non-trivial four-cycle in the D4 background. Such a D4-brane is
realised as a small instanton configuration in the world-volume gauge theory
on the probe D8-brane. The action of the model is:
\beqa
S&=&-\kappa\int~d^4x~dz\Tr\bigg[{1\over2}(1+z^2)^{-1/3}
{\cal{F}}_{\mu\nu}^2\nnb\\
&+&(1+z^2){\cal{F}}_{\mu z}^2\bigg]
+{N_c\over24\pi^2}\int~\omega_5({\cal{A}}),
\enqa
where ${\cal{A}}$ is the 5D $U(N_f)$ gauge field and ${\cal{F}}$ is the field 
strenght. The constant $\kappa$ is related to the 't Hooft coupling, $
\lambda_t$, and $N_c$ by:
\be
\kappa={\lambda_tN_c\over216\pi^3}.
\ee
and $\omega_5({\cal{A}})$ is the CS 5-form for the $U(N_f)$ gauge field 
defined as
\be
\omega_5({\cal{A}})=\Tr\left[{\cal{A}}{\cal{F}}^2-{i\over2}{\cal{A}}^3{\cal{F}}
-{1\over10}{\cal{A}}^5\right].
\ee

Baryon in this model corresponds to a slowly rotating soliton with 
non-trivial instanton number on the four dimensional space parametrized 
by $x^M~(M=1,2,3,z)$.

To check whether this model satisfies Eq.~(\ref{Rre}), we 
use the expressions for the currents, derived in ref.~\cite{HSS}
using the fact that the instantons are localized arbitrarily well at $z=0$ in 
the large $\lambda_t$ limit we get:
\beqa
\label{SS_form_factors}
G_E^{I=0}(r) &=&-\sum_{n=1}^\infty g_{v^n} \psi_{2n-1}(0) Y_{2n-1}(r),\nnb\\
G_M^{I=0}(r) &=& - {9 \pi r \over 4 \lambda N_c  }\sum_{n=1}^\infty g_{v^n} 
\psi_{2n-1}(0)  \nnb\\
&\times& m_{2n-1}Y_{2n-1}(r), \nnb\\
G_E^{I=1}(r) &=&-\sum_{n=1}^\infty g_{v^n}\psi_{2n-1}(0) Y_{2n-1}(r),\nnb\\
G_M^{I=1}(r) &=&-\frac{N_c}{3} \sqrt{\frac{2}{15}} \sum_{n=1}^\infty g_{v^n}
\psi_{2n-1} (0)\nnb\\
&\times&  \rho_{2n-1}Y_{2n-1}(r),
\enqa
where $\left\{\psi_n(z)\right\}$ is a complete set
of functions normalized so that $\psi(z)\sim \kappa^{-1/2}$ that satisfy 
$-(1+z)^{1/3} \partial_z (k(z)\partial_z \psi_n (z) ) = \rho^2_n \psi_n (z)$, 
where the eigenvalues $\rho^{2}_{n}$ (with $\rho_{n+1} >\rho_n$) are related 
to the masses of the vector mesons in this model by $m^{2}_{n} = \rho^{2}_{n} 
M_{KK}^{2}$, the vector meson decay constants $g_{v^n} = 2\kappa\lim_{z\to
\infty} z \psi_{2n-1}(z)$, and $Y_n(r)$ are Yukawa
potentials $Y_{n}(r)=-e^{- \rho_n r}/(4\pi r)$.

From the expressions in Eqs.~(\ref{SS_form_factors}) one can see that $G_E^{I=0}
=G_E^{I=1}$ and that the $r$ dependence of $G_M^{I=0}$ is  the same as 
$G_M^{I=1}$. This is very different from the $r$ dependence
obtained for these form factors in the Skyrme model, as shown in 
ref.~\cite{nos}. As a matter of fact, the relation between $G_E^{I=0}$ and 
$G_E^{I=1}$ was noticed in ref.~\cite{HSS}, since they got the same expression 
for the scalar and isovector charge distributions: 
$\rho_{I=1}(r)=\rho_{I=0}(r)$, although the isovector mean square radius 
should be divergent in the chiral limit. They argue that there is no 
contradiction in this result, 
since the divergence of the isovector mean square radius is due to the IR 
divergence of the pion loop, which is not included in their model. 
However, this result could be an indication that the chiral symmetry breaking
is not correctly incorporated in their model and, in this case, we do not 
expect their form factors to satisfy the relation in Eq.~(\ref{Rre}).
Besides, in the large $r$ limit the scalar form factor is dominated by the
3-pion coupling with the nucleon, whereas the isovector form factor is
dominated by the 2-pion coupling. Therefore, in the large $r$ limit one 
expects a different $r$ dependence for the scalar and isovector form factors.

Taking the large $r$ limit of the form factors in Eqs.~(\ref{SS_form_factors}) 
we find that
\beqa
\label{SS_large_r}
\lim_{r\rightarrow \infty} G_E^{I=0}(r) &= &\frac{g_{v^1} \psi_{1}(0) }
{4\pi r}  e^{- \rho_1 r}, \\
\lim_{r\rightarrow \infty} G_M^{I=0}(r) &= &{9 \pi r \over 16 \pi \lambda 
N_c  } g_{v^1}\psi_1(0) \rho_1 {e^{-\rho_1 r}},\nnb\\
\lim_{r\rightarrow \infty} G_E^{I=1}(r) &= &\frac{g_{v^1} \psi_{1}(0) }
{4\pi r}  e^{- \rho_1 r}, \nnb \\
\lim_{r\rightarrow \infty} G_M^{I=1}(r) &= &\frac{N_c}{12\pi} \sqrt{\frac{2}
{15}}   g_{v^1}\psi_1(0){\rho_1}~{e^{-{\rho_1}r}}\nnb  \;,
\enqa
which makes it easy to see that
\be
\lim_{r \rightarrow \infty} \frac{{G}_E^{I=0} {G}_E^{I=1}}{{G}_M^{I=0} 
{G}_M^{I=1}} = \frac{\lambda_t \sqrt{40/3}  }{\pi \rho^{2}_{1} r^2} \; .
\ee

As expected, the ratio is sensitive to model parameters, so 
Eq.~(\ref{Rre}) is not obeyed and the model does not correctly encode 
the large $N_c$ chiral properties of baryons.  This is troubling in that the 
ability to describe chiral symmetry and its spontaneous breaking are supposed 
to be principal virtues of the model. As mentioned above,
the form factors 
depend only on couplings to vector mesons, and not to pions, in contradiction
 to large $N_c$ $\chi PT$.  This makes the failure of the model to satisfy 
Eq.~(\ref{Rre})  unsurprising.    While the symptoms of the problem are clear, 
whether they represent a technical difficulty in the implementation of the 
model or a deeper structural problem remains an open question.  It appears 
likely that the issue is connected to a non-commutativity of the large 
$\lambda_t$ and chiral limits in this model.

\section{Discussion}

Our analysis above shows that the construction of the
holographic model requires a number of ad hoc assumptions that are not
always consistent  with the low energy regime of QCD. The behavior of model 
independent relations, like the one in Eq.~(\ref{Rre}), is an explicit 
probe of the self-consistency of the assumptions.  Clearly, if one wants to 
match the key features of large $N_c$ QCD in a consistent way, it is essential
 to capture the scale dependence of QCD on the 5D side of the model.  This 
amounts to trying to improve on the ad hoc approximations involved in the
construction of the 5D model.

Although we have focused our analysis on the  models of
refs.~\cite{PomarolWulzer,PW,HSSY,HSS}, the problems in reproducing the 
low-energy
regime of QCD apply rather broadly to  holographic models of QCD.

{\it Acknowledgements.} A.C. and T.D.C. acknowledge the support of the US 
Dept. of Energy.  M.N. acknowledges the support of CNPq and FAPESP.

\end{document}